# Simulation study of intra-beam scattering effect in the HALF storage ring with Piwinski model

**C.W. Luo, P.H. Yang, G.W. Liu, W.W. Li, N. Hu, W.M. Li, Z.H. Bai, L. Wang**

*National Synchrotron Radiation Laboratory, University of Science and Technology of China,*
Hefei 230029, China
*E-mail*: wanglin@ustc.edu.cn

ABSTRACT: The Hefei Advanced Light Facility (HALF) will be a VUV and soft X-ray diffraction-limited storage ring (DLSR), and its high density of electron bunches makes the intra-beam scattering (IBS) effect very serious. In this paper, an IBS module used in the IMPACT code is developed, where the scattering process of IBS is described by the Piwinski model in Monte Carlo sampling. For benchmarking, the IMPACT code with IBS module is compared with the ELEGANT code and a semi-analytic code using Bane's model. Then, the results of IBS effect in the HALF storage ring studied by this new code are presented. With various countermeasures, the IBS impact can be controlled to a certain extent, and the expected beam emittance is approximately 59 pm·rad.

# Contents

## 1. Introduction

In recent decades, increasing the brightness of synchrotron radiation has been one main motivation for the development of storage ring light sources [1]. The primary way is to reduce the electron beam emittance of the storage ring. A light source called diffraction-limited storage ring (DLSR) with electron beam emittance in both horizontal and vertical planes comparable to the diffraction-limited photon emittance has become the main development trend, which can produce synchrotron radiation with very high brilliance and good transverse coherence [2]. To reduce the emittance down to this level, multi-bend achromat (MBA) lattices are adopted in DLSR designs [3]-[4], such as MAX-IV [5], Sirius [6], ESRF-EBS [7], and ALS-U [8]. The ultra-low emittance can contribute to very high beam intensity and thus serious intra-beam scattering (IBS), which is caused by small angle Coulomb scattering between charged particles in a bunch and will increase equilibrium emittance and energy spread [9]-[10].

The IBS theory was first introduced by Piwinski [11], later extended by Martini, and their outcome is called the standard Piwinski method [12]. During the same period, Bjorken and Mtingwa used a different approach to describe IBS named B-M formula, taking into account the strong focusing effect [13]. Karl Bane developed the modified Piwinski method that includes the strong focusing effect [14], and an approximation formula to describe IBS at high beam energy [15]. The Piwinski method was also developed as completely integrated modified Piwinski (CIMP) formula by Kubo to simplify numerical calculation [16].

These analytical models that describe the IBS effect adopt the assumption of Gaussian beam distribution. However, the beam distribution is usually non-Gaussian in the real machines, especially for proton and ion accelerators [17]. Therefore, it is important to develop numerical simulation tools to study the IBS effect of non-Gaussian beams together with other diffusion mechanisms, such as wake field. MOCAC [18] and BETACOOL [19] are well-known numerical simulation codes to study the IBS effect in hadron accelerators, where IBS is very serious due to lack of radiation damping mechanism. With the development of electron accelerators with very low beam emittance, a new code, named SIRE [20], was developed, where the equilibrium beam emittance is calculated numerically by emittance growth equations. The emittance growth rates induced by IBS in SIRE are obtained by Monte Carlo sampling instead of analytic formulas.

To include more realistic motion of electrons in the storage ring to study the IBS effect, in this paper we will develop a simulation code based on the PIC/MCC method [21], in which the PIC (particle in cell) process will be used to describe electron motion under external electromagnetic field and the MCC (Monte Carlo collision) process will be used to simulate the Coulomb scattering in a bunch. The IMPACT [22] code will be used as the primary tool to track electrons and obtain emittance evolution of electrons for the PIC process, and Piwinski model will be applied in the Monte Carlo sampling for the MCC process. The developed code will be benchmarked with other existing methods and used to study the IBS effect in the Hefei Advanced Light Facility (HALF) storage ring.

The paper is organized as follows. In Section 2, the main parameters of the HALF storage ring are described. In Section 3, the basic theory and method to simulate the IBS effect are introduced, and the developed new code based on Piwinski model is described and then benchmarked. Then in Section 4, the new code is applied to the HALF storage ring, and the influence and suppression of IBS in HALF is studied in detail. Finally, conclusions are given in Section 5.

## 2. The HALF storage ring

HALF is a VUV and soft X-ray DLSR [23]-[25] with a beam energy of 2.2 GeV and a natural emittance aiming to be lower than 100 pm·rad, which corresponds to the diffraction-limited emittance of photon beam with energy of 1~2 keV. Except for the requirements for ultra-low emittance, good nonlinear dynamics performance and more number of straight sections are the main design and optimization concerns of the HALF storage ring. A new-type modified hybrid 6BA lattice with longitudinal gradient bends and reverse bends as well as a short straight section in the middle part was put forward and optimized for the HALF storage ring [26]. The optical functions of a lattice cell are plotted in figure 1. The main parameters of the storage ring are listed in Table 1. The natural beam emittance is approximately 86 pm·rad. The betatron tunes are (48.15, 17.15), on the difference resonance line, for performing full-coupling beam operation, which can reduce the horizontal emittance, suppress the IBS effect and increase beam lifetime [26]. The horizontal and vertical zero-current beam emittances are 50.2 pm·rad under the full coupling condition [27]:

$$j_x \varepsilon = j_x \varepsilon_x + j_y \varepsilon_y, \quad \varepsilon_y = \kappa \varepsilon_x, \tag{1}$$

where $\varepsilon$ is the natural beam emittance, $\varepsilon_{x,y}$ are the horizontal and vertical zero-current beam emittances, respectively, $j_{x,y}$ are the horizontal and vertical damping partition numbers, respectively, and $\kappa$ is the transverse coupling ratio.

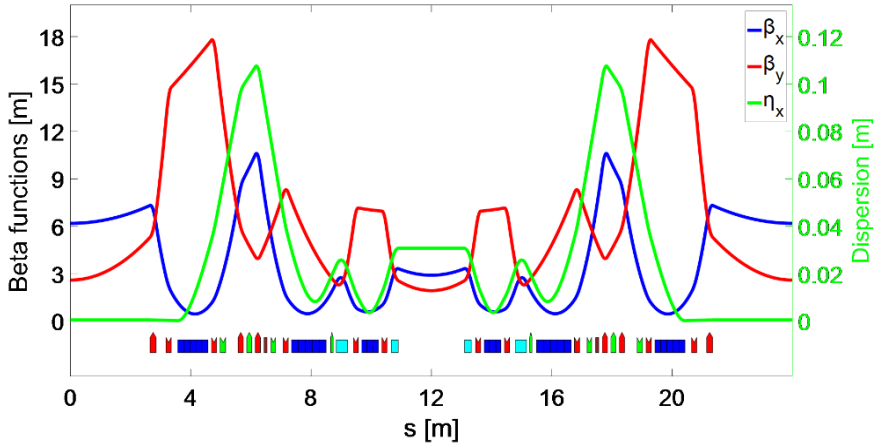

**Figure 1.** Linear optical functions and magnet layout of the HALF 6BA lattice. In the magnet layout, bends are in blue blocks, reverse bends in cyan, quadrupoles in red, sextupoles in green and octupoles in brown.

Table 2. Main parameters of the HALF storage ring.

| Parameter | Value | Units |
|---|---|---|
| Beam energy | 2.2 | GeV |
| Circumference | 479.86 | m |
| Number of cells | 20 | |
| Natural emittance | 86.3 | pm·rad |
| Energy loss per turn | 186.7 | keV |
| Natural energy spread | 6.2 | $10^{-4}$ |
| Momentum compaction factor | 9.0 | $10^{-5}$ |
| Damping partition number (H/V/E) | 1.39/1/1.61 | |
| Natural damping times (H/V/E) | 27.2/37.7/23.4 | ms |
| Betatron tunes (H/V) | 48.15/17.15 | |
| RF frequency | 499.8 | MHz |
| Harmonic number | 800 | |

Low beam emittance, weak radiation damping, high bunch charge and short bunch length are important factors that make the IBS effects serious in the HALF storage ring, so it is necessary to study the IBS effects extensively.

## 3. IBS simulation with Piwinski model

### 3.1 Theoretical basis

When electrons are moving in a storage ring, electrons will be forced by external electromagnetic field and continuously scattered by other electrons. The motion process of electrons can be described by Boltzmann equation, as shown in [21]. For small time-step, the equation can be linearized and solved at any time by step-by-step calculations starting from an initial velocity distribution function. Therefore, during one step, there are two discrete procedures to decouple the non-collisional and collisional parts of the motion: first obtaining $f^* = (1 + \Delta t \boldsymbol{D})f$ and then obtaining $(1 + \Delta t \boldsymbol{J})f^*$ for a cell located at $\boldsymbol{r}$. Here, $f^*$ and $f$ are the velocity distribution functions, $\boldsymbol{D}$ and $\boldsymbol{J}$ denote noncollisional and collisional operators, respectively, $\Delta t$ is the time interval, $\boldsymbol{r}$ is the spatial vector. For non-collisional parts, the motion is governed by the Lorentz force $\boldsymbol{F} = q(\boldsymbol{E} + \boldsymbol{v} \times \boldsymbol{B})$, and the equations are shown as follows,

$$m\frac{d\boldsymbol{v}}{dt} = q(\boldsymbol{E} + \boldsymbol{v} \times \boldsymbol{B}), \boldsymbol{v} = d\boldsymbol{r}/dt, \tag{2}$$

where $m$ is the mass of electron, $\boldsymbol{v}$ is the velocity vector, t is the time, q is the charge of electron, $\boldsymbol{E}$ and $\boldsymbol{B}$ are the electric and magnetic fields, respectively. For collision parts, the velocity distribution function $f(\boldsymbol{v},\boldsymbol{r},t)$ after scattering is given by

$$f(\boldsymbol{v},\boldsymbol{r},t+\Delta t) = \frac{n}{N}\sum_{i=1}^{N}[(1-P_i)\delta^3(\boldsymbol{v}-\boldsymbol{v}_i) + P_i Q_i(\boldsymbol{v})], \qquad (3)$$

where $n$ is the number density for the cell, $N$ is the number of simulated electrons in the cell, $\delta^3()$ is the delta function, $P_i$ is the collision probability, and $Q_i(\boldsymbol{v})$ is the probability density function for the post-collision velocity.

Based on the above theoretical framework, PIC/MCC methods have been proposed and applied in various simulations, especially in the plasma physics region. Here, we will employ the PIC/MCC method to study the IBS effects in electron storage rings. The electron beam is modeled by macroparticle, and by tracking the movement behaviors of each macroparticle, the performance of the whole beam can be obtained statistically.

For a storage ring, the symplectic property in numerical algorithm is crucial for long-term tracking. Here, the IMPACT code is selected as the primary tool due to its symplectic property in numerical simulation and parallel computing capability, where many external electromagnetic elements have been modeled, such as dipole, quadrupole, sextupole, RF cavity, etc. For the electron storage ring simulation in the IMPACT code, there is a lack of radiation damping process. We will implement the radiation damping and quantum excitation model in the code, and the algorithm of the model is [28]:

$$\begin{cases} \tilde{\delta}_{i+1} = \delta_{i+1}(1-D_p) + \sigma_p\sqrt{2D_p}\cdot\delta_{rand} \\ \tilde{x}_{i+1} = x_{i+1} + \sigma_x\sqrt{D_x}\cdot x_{rand} \\ \tilde{x}'_{i+1} = x'_{i+1}(1-D_x) + \sigma_{x'}\sqrt{D_x}\cdot x'_{rand} \, , \\ \tilde{y}_{i+1} = y_{i+1} + \sigma_y\sqrt{D_y}\cdot y_{rand} \\ \tilde{y}'_{i+1} = y'_{i+1}(1-D_y) + \sigma_{y'}\sqrt{D_y}\cdot y'_{rand} \end{cases} \qquad (4)$$

with the coefficients $D_{p,x,y} = j_{p,x,y}U_0/E_0$, where $j_{p,x,y}$ are the longitudinal and transverse damping partition numbers, $U_0$ is the energy loss per turn, $E_0$ is the beam energy, $\sigma_p$ is the energy spread, $\sigma_{x,y}$ are the transverse beam sizes and $\sigma_{x',y'}$ are the transverse angles of divergence, $\delta_{rand}, x_{rand}, x'_{rand}, y_{rand}$, and $y'_{rand}$ are random numbers drawn from normal distributions with unit standard deviation.

### 3.2 Piwinski model

The sampling method is based on the Piwinski model [11]. The main process of intra-beam scattering is modeled by a two-body collision model. In the center-of-mass (CM) coordinate system, as shown in figure 2, the velocity vector after collision can be easily obtained by conservation law of energy and momentum. Using the Piwinski model, in the laboratory coordinate system, the changes in momentum for particles 1 and 2 can be expressed as follows,

$$\Delta P_{1,2x} = \pm\frac{P}{2}\left[\left(\zeta\sqrt{1+\frac{\xi^2}{4\alpha^2}}\sin\phi - \frac{\xi\theta}{2\alpha}\cos\phi\right)\sin\varphi + \theta(\cos\varphi - 1)\right], \qquad (5)$$

$$\Delta P_{1,2y} = \pm\frac{P}{2}\left[\left(\theta\sqrt{1+\frac{\xi^2}{4\alpha^2}}\sin\phi - \frac{\xi\zeta}{2\alpha}\cos\phi\right)\sin\varphi + \zeta(\cos\varphi - 1)\right], \qquad (6)$$

$$\Delta P_{1,2s} = \pm\frac{P}{2}[2\alpha\gamma\cos\phi\sin\varphi + \gamma\xi(\cos\varphi - 1)], \qquad (7)$$

where $\varphi$ and $\phi$ are the polar angle and azimuthal angle, $\xi = \frac{P_1-P_2}{\gamma P}$, $\theta = x'_1 - x'_2$, $\zeta = y'_1 - y'_2$, $\alpha = \frac{\sqrt{\theta^2+\zeta^2}}{2}$, $P_1, P_2, x'_1, x'_2, y'_1$ and $y'_2$ are the momentum and the slope of velocity in horizontal and vertical directions of the two particles, respectively, P is the mean momentum of the two particles, and γ is the Lorentz factor. There are two angles to be determined in the calculation. The first is the equivalent polar angle, namely the scattering angle, which is defined as [29]:

$$\varphi = \frac{2r_0}{\gamma}\sqrt{\frac{\pi c n \Delta t_s L_c}{\bar{\beta}^3}}, \tag{8}$$

where $r_0$ is the classical radius of electron, $c$ is the speed of light, $n$ is the electron density in a cell, $L_c$ is the Coulomb logarithm, $\bar{\beta}$ is the average speed of the two particles, and $\Delta t_s = L_{\text{element}}/c$ is the time step for the macroparticle movement in the simulation. Additionally, the azimuthal angle $\phi$ is theoretically distributed uniformly in $[0, 2\pi]$ and obtained by Monte Carlo sampling.

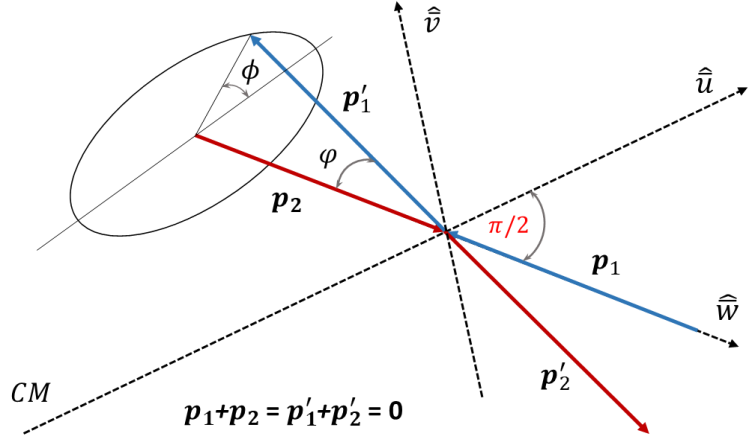

**Figure 2.** Changes of momentum in a two-particle collision in the CM coordinate system. $(\hat{u}, \hat{w}, \hat{v})$ is the velocity unit of the CM frame, $\varphi$ is the polar angle and $\phi$ is the azimuthal angle.

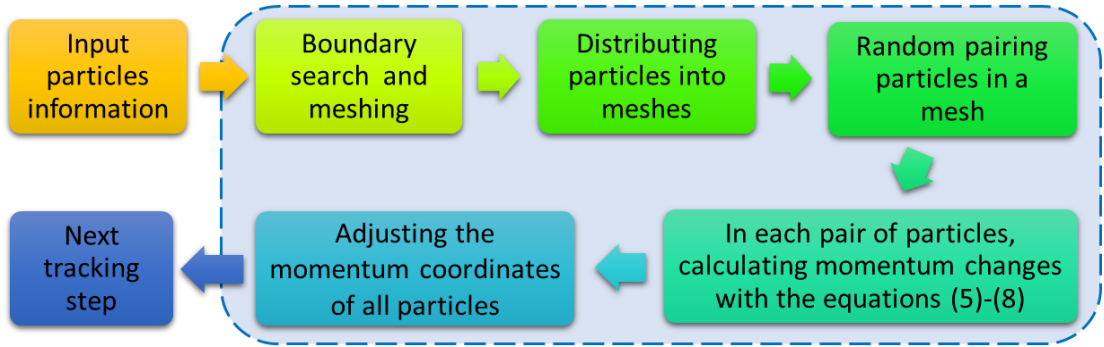

**Figure 3.** Procedure of the IBS module.

Based on the binary collision model using the Piwinski method, we construct an IBS module describing the scattering process. In this module, the first step is phase space meshing, including transverse and longitudinal planes and considering local beam size information. Then, the macroparticles are paired according to their positions. Using formulas (5)-(8), the momentum vectors of macroparticles are obtained. Therefore, the motions of macroparticles in the storage ring are imitated in several steps, as illustrated in figure 3.

In the simulation with the IMPACT code with IBS module, after the parameters initialization, the lattice elements are split into many thin pieces. The PIC processes inside the thin pieces are handled by the original elements of the IMPACT code [22][30]. Between thin pieces, the scattering events occur and are processed by the IBS module. After one turn revolution of the storage ring, the effects of radiation damping and quantum excitation are added. The transverse and longitudinal phase space coordinates of each macroparticle at specified locations are recorded every turn. The global beam parameters, such as beam emittance, bunch length and beam energy spread, can be obtained by statistical analysis of recorded data. For the IBS effect, it is usually necessary to track for about tens or hundreds of milliseconds to achieve the equilibrium state. In the following study of IBS effect, the final equilibrium state of electron beam is our concern and can be obtained from the IMPACT code with IBS module.

### 3.3 Benchmarking of the IMPACT code with IBS module

We will use the ELEGANT code [31] and a semi-analytic code developed in our previous work [32] to benchmark the IMPACT code with IBS module. The ELEGANT code is widely-used accelerator physics design code in the accelerator community, where a specialized command tool exists to calculate the IBS effect [31]. In our previous work [32], we obtained the equilibrium beam emittance and energy spread by numerically solving emittance growth equations, in which the emittance growth rates are obtained from Bane's high energy approximation formulas.

Using the HALF storage ring with full transverse coupling as an example, these codes will be used to study the IBS effects. With the same initial beam distribution parameters, we calculated the horizontal equilibrium beam emittance and energy spread as a function of bunch charge, and the results are shown in figure 4. Here, the bunch is not lengthened and damping wiggler is not used. The ELEGANT code gives the strongest emittance growth and energy spread increase while the semi-analytic code gives the weakest, and the results from the IMPACT code with IBS module are between them. In the IBS calculation of ELEGANT, the calculated growth rates of emittance and energy spread are based on B-M formula, derived from the Gaussian beam distribution assumption and simplification [13][31]. In the semi-analytic code [32], when solving equilibrium emittance equations to obtain the growth rates of emittance and energy spread, the overall average parameters of storage ring are used and the parameter variations along the storage ring are ignored. The discrepancies between their results are approximately 10% level in emittance and within 5% level in energy spread. The discrepancies are acceptable and in the following we will use the IMPACT code with IBS module to study the IBS effect in the HALF storage ring.

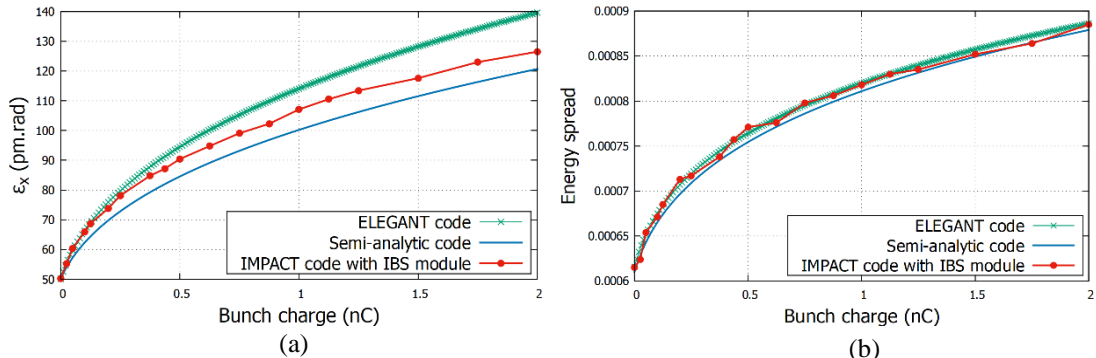

**Figure 4.** The variations of equilibrium emittance (a) and energy spread (b) with bunch charge.

## 4. Application to the HALF storage ring

As shown in figure 4, although with full transverse coupling, the equilibrium beam emittance is nearly twice the zero-current emittance at the bunch charge of 0.8 nC. Therefore, it is necessary to employ other countermeasures and assess their effectiveness.

## 4.1 Bunch lengthening

Bunch lengthening using higher harmonic cavity is an effective countermeasure to suppress emittance growth due to the IBS effect. This is because an increase in bunch length leads to a corresponding increase in the volume of the bunch, which results in a decrease in particle density. In our code, it is convenient to adjust the bunch length with the harmonic RF cavity element, and the macroparticle simulation also indicates that bunch lengthening can be easily got. We simulated the IBS effect including the evolution of beam emittance and energy spread with bunch lengthening and bunch charge. The full coupling beam was also used in the simulation. Table 2 shows the simulation results.

**Table 2.** Horizontal equilibrium emittances and energy spreads for different bunch lengthening factors and bunch charges.

| Bunch charge (nC) | Horizontal emittance $\varepsilon_{x,IBS}$ (pm·rad) | | | Energy spread $\sigma_{p,IBS}$ ($10^{-4}$) | | |
|---|---|---|---|---|---|---|
| | Without lengthening | 3 times lengthening | 5 times lengthening | without lengthening | 3 times lengthening | 5 times lengthening |
| 0.4375 | 84.5 | 67.5 | 62.3 | 7.57 | 6.83 | 6.70 |
| 0.875 | 102.3 | 77.5 | 71.6 | 8.06 | 7.21 | 6.98 |

At the bunch charge of 0.875 nC, corresponding to 400 mA beam current with about 90% buckets equally filled, for bunch lengthening factors of 3 and 5, the equilibrium beam emittances are reduced by 24.2% and 30%, and the energy spreads are reduced by 10.5% and 13.4%, respectively, compared to the case without lengthening. At the bunch charge of 0.4375 nC, the equilibrium beam emittances are reduced by 20.1% and 26.3% for bunch lengthening factors of 3 and 5. For the bunch lengthening cases, the beam equilibrium emittances are lower than the emittance goal of the HALF storage ring.

## 4.2 Damping wiggler

We will study another countermeasure, using damping wiggler (DW). At present, the DW model is not completed in IMPACT. So we will use another method to test the effectiveness of DW. With the use of DW, the natural emittance, energy spread and radiation damping time will be changed. Their analytic expressions are given by [33]-[34]:

$$\frac{\varepsilon_W}{\varepsilon_0} = \frac{1+\frac{2\sqrt{2}\pi^3 \rho_0^3 K^3}{\lambda_W^3 \gamma^3 \langle H_0 \rangle}\langle H_W \rangle}{1+\frac{\pi N \rho_0 K^2}{\lambda_W \gamma^2}}, \tag{9}$$

$$\frac{\sigma_{p_W}^2}{\sigma_{p_0}^2} = \frac{1+\frac{\sqrt{2}\pi^2 N \rho_0^2 K^3}{\lambda_W^2 \gamma^3}}{1+\frac{\pi N \rho_0 K^2}{\lambda_W \gamma^2}}, \tag{10}$$

$$\tau_i = \frac{2E_0 T_0}{j_i(U_0+U_W)}, i = x, y, p, \ U_W = \frac{8\pi^2 r_0 mc^2 \gamma^2 K^2}{3\lambda_W^2} N\lambda_W, \tag{11}$$

where $\rho_0$ is the radius of bending magnet, K=0.934 B [T] λw[cm], B is the magnetic field of DW, $\lambda_w$ is the wiggler period length and $N$ is the number of wiggler periods, $\langle H_{w,\,0} \rangle = \frac{1}{C}\oint \frac{ds}{\beta_w}\left[\eta^2 + \left(\beta_w \eta' - \frac{1}{2}\beta_w' \eta\right)^2\right]$, C is the storage ring circumference, $\beta_w$ and $\beta_w'$ are the average horizontal beta function and its derivation in the wiggler, $\eta$ and $\eta'$ are the dispersion function and its derivation. And $E_0$ is the beam energy, $T_0$ is the revolution time of storage ring, $U_0$ is the energy loss per turn without DW, $U_W$ is the energy loss per turn with DWs.

After obtaining the natural emittance, energy spread and damping times with DWs, we can change the parameters of the radiation damping and quantum excitation module to simulate the IBS process. Therefore, the effectiveness of DW can be analyzed quantitatively. In the physics design of the HALF storage ring, two straight sections are used to install DWs, which have length of about 4.2 m. Generally, short period is beneficial for reducing emittance and high peak magnetic field helps to enhance the radiation damping. However, considering the technology limitations, it is hard to simultaneously get both very short period and very high field. Therefore, we optimized the period length and peak magnetic field of DW to obtain a lower equilibrium emittance. With different DW parameters, the natural damping time, energy spread and emittance were calculated for HALF, which are listed in Table 3. It indicates that after using DWs, the natural damping times were reduced by 30% ~ 36%, which can help to reduce emittance growth due to the IBS effect more effectively. The natural emittance reduction and natural energy spread increase are also beneficial for obtaining lower equilibrium beam emittance.

The IBS simulation results of horizontal equilibrium emittance are also listed in Table 3, where the bunch charge is 0.875 nC and there is no bunch lengthening. The equilibrium emittances with 100% coupling are reduced by 13~16% compared to those without DWs. According to these results and other comprehensive concerns, the DW with peak field of about 1.9 T and period length of about 100 mm is preferred.

**Table 3.** Main parameters and simulation results for DWs with different period lengths and peak magnetic field strengths.

| Parameter | DW1 | DW2 | DW3 | DW4 | DW5 | DW6 |
|---|---|---|---|---|---|---|
| $B_{peak}$(T)/$\lambda_{DW}$(mm) | 1.75/75 | 1.80/80 | 1.85/90 | 1.90/100 | 1.95/110 | 2.0/120 |
| $U$ (keV) | 265.5 | 270.3 | 273.5 | 279.6 | 284.1 | 292.5 |
| $\tau_y$ (ms) | 26.5 | 26.1 | 25.8 | 25.2 | 24.8 | 24.1 |
| $J_x$ | 1.274 | 1.269 | 1.266 | 1.260 | 1.256 | 1.248 |
| $\sigma_\delta$ ($\times 10^{-4}$) | 7.04 | 7.12 | 7.20 | 7.29 | 7.38 | 7.49 |
| $\varepsilon_{nat,DW}$ (pm·rad) | 69.7 | 70.0 | 71.1 | 72.6 | 74.9 | 77.9 |
| $\varepsilon_{x,DW}$ (pm·rad) (10% coupling) | 271 | 259 | 257 | 254 | 258 | 259 |
| $\varepsilon_{x,DW}$ (pm·rad) (100% coupling) | 89.1 | 87.9 | 87.6 | 86 | 86.8 | 88.3 |

### 4.3 Bunch lengthening and damping wiggler

Simultaneously using DW and bunch lengthening would be more helpful for suppressing the IBS effects. In the simulation for HALF, two 4.2 m long DWs with peak magnetic field of 1.9 T and period length of 100 mm were used, and the bunch charge was 0.875 nC. The equilibrium emittances were 64.5 pm·rad and 59.3 pm·rad for bunch lengthening factors of 3 and 5, respectively, further reduced by 16.8% and 17.2% compared to the case without DW. The variations of horizontal equilibrium beam emittance and energy spread with bunch charge are shown in figure 5.

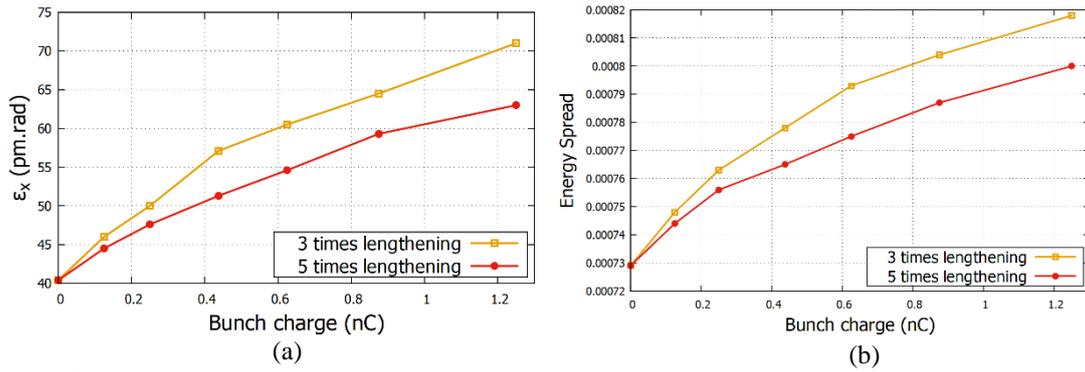

**Figure 5.** The variations of horizontal equilibrium emittance (a) and energy spread (b) with bunch charge for different bunch lengthening cases.

### 4.4 Beam energy

Although harmonic cavity and DW are effective countermeasures to resist the emittance growth due to the IBS effect, it still shows non-negligible effects. The main reasons are ultra-low emittance, weak radiation damping and relatively low beam energy, which result in large emittance growth rates. The natural beam emittance is proportional to the square of beam energy, but the IBS emittance growth rate is negatively correlated with beam energy. From the emittance reduction point of view, there will be an optimal beam energy. Then, we scanned the beam energy for the HALF storage ring to study its effect on IBS. In the simulation, the natural emittance, radiation damping and other parameters will be changed with beam energy. Figure 6 shows the variation of horizontal equilibrium emittance with beam energy considering the IBS effect. The bunch charge is 0.875 nC. The bunch length was stretched by a factor of 3 and 5, and the full coupling beam was used. Two 4.2 m long DWs with peak magnetic field of 1.9 T and period length of 100 mm were adopted. If there is no bunch lengthening and DW, the equilibrium beam emittance has the minimum value at about 2.7 GeV. After using the countermeasures, the optimal beam energy is about 2.4 GeV, and the equilibrium emittance is around 60 pm·rad. Therefore, we can promote the beam energy from the designed 2.2 GeV to 2.4 GeV in order to reduce the IBS effect and obtain lower equilibrium emittance. Besides, a higher energy is also beneficial for suppressing beam instabilities and increasing radiation flux. At present, HALF has considered increasing beam energy in the future.

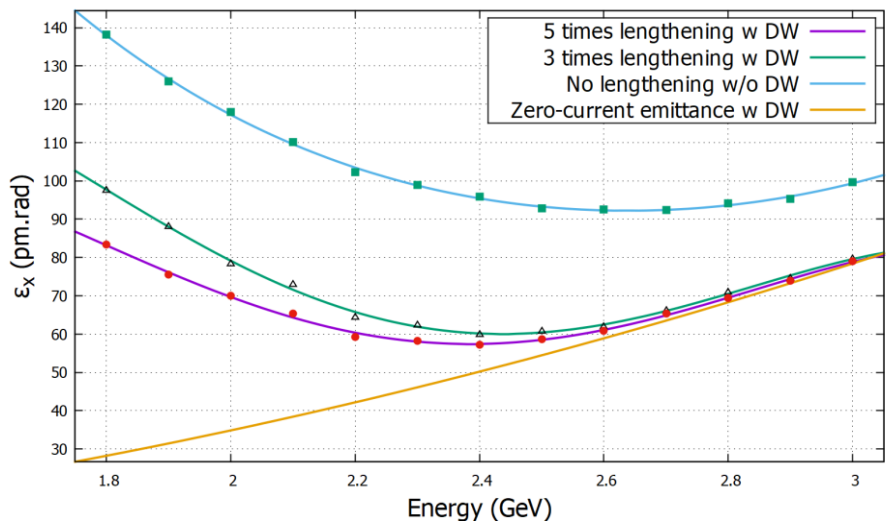

**Figure 6.** The variation of horizontal equilibrium emittance with beam energy.

## 5. Conclusion and outlook

To study the IBS effect more accurately, we have developed a simulation code based on PIC/MCC methods, and the Piwinski model was used to simulate the IBS process. It was added into the IMPACT code as a new module, named IBS module. The simulation results agree well with those of the ELEGANT code and the semi-analytic calculation of Bane's model. The new simulation tool provides a new option of the IBS study for HALF and other electron storage rings.

The IBS effects of the HALF storage ring were studied detailedly by the IMPACT code with IBS module. Although the emittance growth due to IBS is very severe, the equilibrium beam emittance can be controlled to be about 59 pm·rad in the horizontal and vertical planes for the current 400 mA with various countermeasures, including full transverse coupling, bunch lengthening and DWs. To reduce the IBS effect and lower equilibrium emittance further, we will consider increasing the beam energy of HALF in the future.

In the future, we will improve the new simulation tool further. On one hand, the lattice model constructed in IMPACT will be more complete to include other components. On the other hand, the IBS module will be upgraded to enhance its computing efficiency and incorporate more scattering processes.

## Acknowledgments

This work was supported by the HALF project and the National Key R&D Program Project (2016Y-FA0402000).